\documentclass[onecolumn,prc,showpacs,a4]{revtex4}
\usepackage{epsfig,graphicx,float,pst-all}
\usepackage{amssymb}
\usepackage{mathrsfs}
\usepackage{amsmath}
\usepackage{url}
\usepackage{esvect}
\usepackage{subfigure}
\usepackage[T1]{fontenc}
\usepackage{textcomp}
\usepackage{gensymb}
\usepackage{url}
\usepackage{multirow,rotating}
\usepackage{setspace}
\usepackage{multind}

\newcommand{\be}{\begin{equation}} 
\newcommand{\ee}{\end{equation}}
\newcommand{\bea}{\begin{eqnarray}} 
\newcommand{\eea}{\end{eqnarray}}
\newcommand{\bc}{\begin{center}}
\newcommand{\ec}{\end{center}}

\begin{document}

\title{ 
New experiments demand for a more precise analysis of continuum spectrum in $^6$He: technical details and formalism}
\author{Jagjit Singh$^1$, L. Fortunato$^1$}
\affiliation{ 1) Dipartimento di Fisica e Astronomia ``G.Galilei''  and INFN-Sezione di Padova,\\ via Marzolo 8, 
I-35131 Padova, Italy.}
\email{jsingh@pd.infn.it}

\begin{abstract}
A simple three-body model of $^6$He is extended to include \textit{sd}-continuum states in the picture in addition to the already investigated \textit{p}-states. The role of different continuum components in the weakly bound nucleus $^6$He is studied by coupling 
unbound spd-waves of $^5$He by using simple pairing contact-delta interaction. The main focus of this paper is to outline the procedure that allows the calculation of different configurations of $^6$He including continuum
states and to set up basic ingredients for computations. The method and results discussed here will be used for the calculation of monopole, dipole, quadrupole and octupole response of $^6$He.
\end{abstract}
\pacs{21.10.Gv, 21.10.Ky, 26.60.Cs}
\maketitle

\section{Introduction}
Motivated by the recent experimental measurements at GANIL \cite{Moug} and in other laboratories \cite{Povo} on continuum resonances in $^6$He, 
we have developed a simple model \cite{Fort} to study the weakly bound ground state and low lying continuum states of 
$^6$He by coupling two unbound \textit{p}-waves of $^5$He. Recently we have extended the model space with 
inclusion of \textit{sd}-continuum waves of $^5$He \cite{JSi, JST}. The large basis set of these \textit{spd}-continuum wavefunctions are used 
to construct the two-particle $^6$He ground state $0^+$ emerging from five different possible configurations
i.e. $(s_{1/2})^2$, $(p_{1/2})^2$, $(p_{3/2})^2$, $(d_{3/2})^2$ and $(d_{5/2})^2$. The simple pairing 
contact-delta interaction is used and pairing strength is adjusted to reproduce the bound ground state of 
$^6$He. Preliminary results shows how the ground state displays collective nature by taking contribution from five
different oscillating continuum states that sum up to give an exponentially decaying bound wavefunction
\cite{JSi}. We have also studied several properties of this state: the mean square distance between the valence nucleons
and the mean square distance of their centre of mass w.r.t core \cite{JSi}.
In the present paper, rather than concentrating on the physics or on the results (that will be presented somwhere else), we will describe the computational procedure in detail and we will discuss some coefficients that are needed in the 
computations.

\section{Procedure}
\begin{figure}[!t]
\epsfig{file=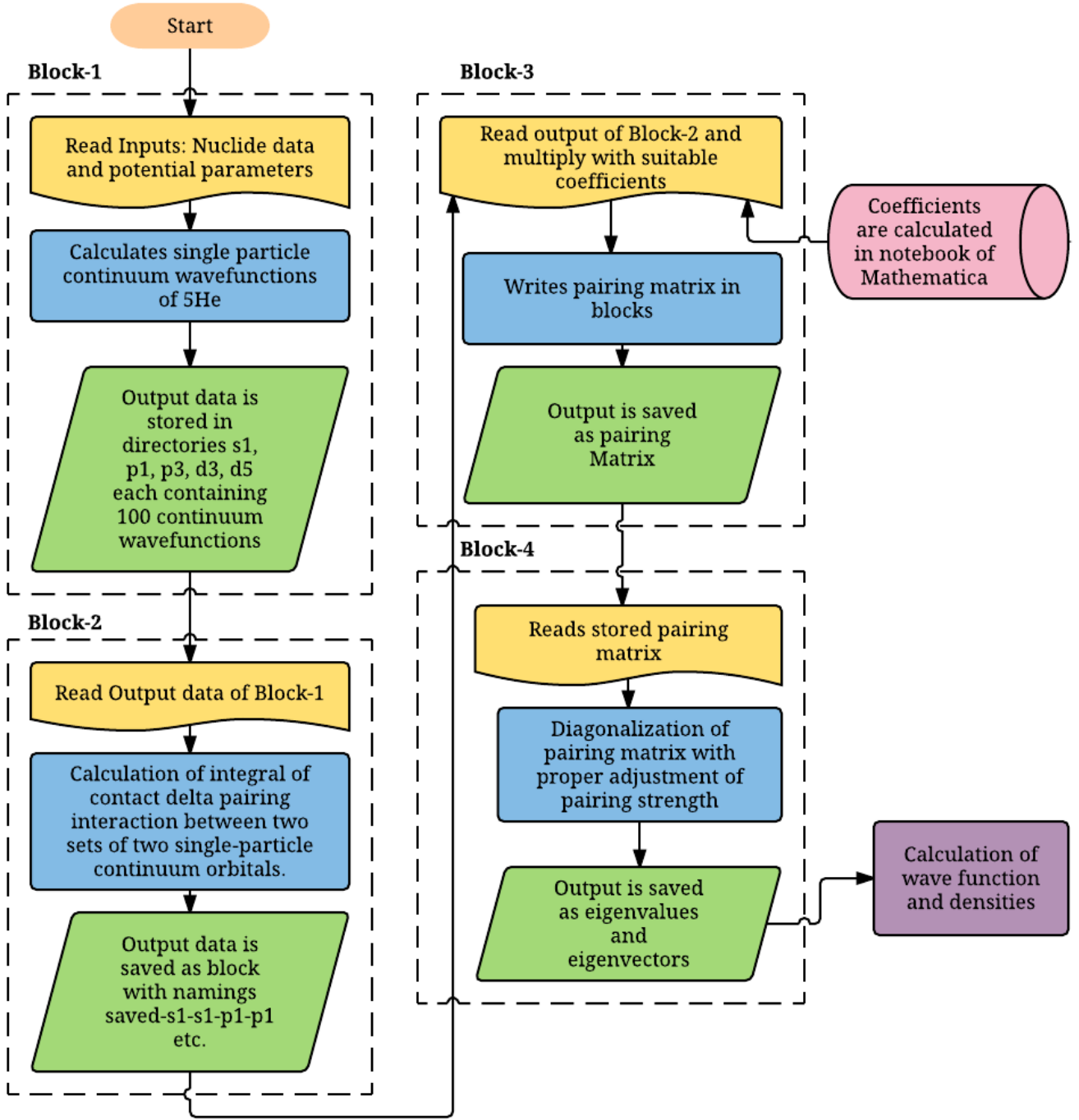, width=0.99\textwidth, clip=}
\caption{(Color online) Flow chart diagram of procedure followed with series of codes used. Blocks 
are indicated with dashed black boxes. Inside each block data reading is indicated as
light yellow cards, algorithms are indicated as blue rectangles and outputs in green.}
\label{Pro}
\end{figure}
The procedure adopted for these calculations is explained in Fig. (\ref{Pro}) with the help of a flow
chart diagram. It is divided in blocks that correspond to the various codes used. In each block yellow background indicates input lines where data must be passed to the code.
Several types of inputs are required: most of them are integers or real number (mostly in convenient nuclear units) that must 
be set case-by-case, others are strings of text. Most input data are read in the input file.
\subsection{Block 1}
\begin{figure}[!t]
\epsfig{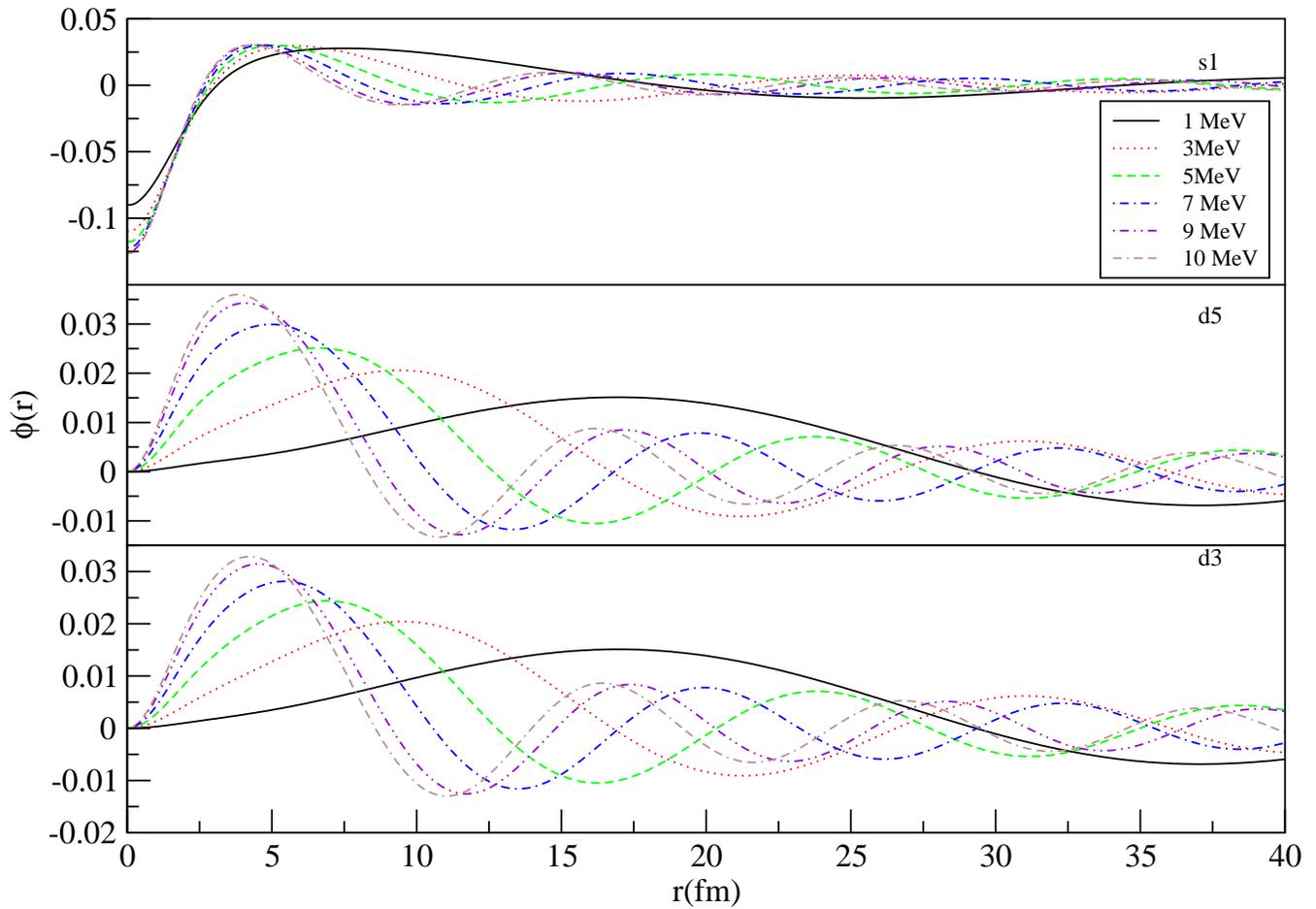}
\caption{(Color online) $^5$He sd-continuum waves as a function of radial variable for continuum energies 1, 3, 5, 7, 9 and 
10 MeV. Each panel is labeled by \textit{l} and $2$\textit{j}.}
\label{5He}
\end{figure}

Block $1$ calculates the $spd-$continuum single-particle states ($E_C>0$) of $^5$He with with Woods-Saxon potential $+$ spin-orbit potential \cite{JSi}, with energies from $0.1$ to $10.0$ MeV on a radial grid that 
goes from $0.1$ to $100.0$ fm (notice that this amount to $2.2$ Mb of data for each component). Examples of these wave functions are shown in Fig. (\ref{5He}), where the $s_{1/2}$, $d_{5/2}$ and $d_{3/2}$ oscillating 
continuum waves are displayed as a function of r in the range of $0-40$ fm for continuum energies $1, 3, 5, 7, 9$ and $10$ MeV.
\subsection{Block 2}
By using the midpoint method as a discretization recipe, the wave functions are normalized to a Dirac delta in energy with an energy spacing of $2.0, 1.0, 0.5, 0.2,$ and $0.1$ MeV corresponding to block basis 
dimensions of N $=5, 10, 20, 50,$ and $100$ respectively. Two-particle states are constructed with proper couplings to J$=0^+, 1^-, 2^+, 3^-$. Then integral of contact delta pairing interaction between two sets of 
two single-particle continuum orbitals are calculated. The output is saved in matrix blocks. For ground state J$=0^+$ and J$=1^-$, it amounts to calculation of $15$ matrix blocks (approximately $\sim 9$ Gb of data) , for J$=2^+$ state 
it amounts to calculation of $28$ matrix blocks, which is a hard computational task and for J$=3^-$, it amounts to calculation of $6$ matrix blocks.
\subsection{Block 3}
It simply reads the blocks evaluated in previous block and multiply them with appropriate coefficients which are evaluated in separate mathematica notebook and calculates the matrix elements of pairing matrix.
The output pairing matrix is huge data set file ($\sim 21$ Gb of data for the largest case in ground state).
\subsection{Block 4}
It simply diagonalizes the pairing matrix with standard routines to give eigenvalues and eigenvectors. The
coefficient of the $δ\delta-$contact matrix, G, has been adjusted to reproduce the correct ground state energy each time. The actual
pairing interaction g is obtained by correcting with a factor that depends on the aforementioned spacing between energy states \cite{JST}.
The biggest adopted basis size gives a fairly dense continuum in the region of interest.
The output of block $4$ is further used for calculation of ground state wavefunction, transition probabilities and two particle densities.

\section{$^6$He wavefunctions and matrix elements of pairing interaction}

The simple model \cite{Fort} with two non interacting particles in the above single particle levels of $^5$He produces different parity states (see Table 1 of \cite{JSi}). 
The two-particle wave functions are constructed by tensor coupling of two continuum states of $^5$He.
The five states of $^5$He are not discrete, but rather depend on the energies of the continuum orbitals. 
Each single particle continuum wavefunction is given by
\begin{equation}
\phi_{\ell,j,m}(\vec{r},E_C)=\phi_{\ell,j}(r,E_C)[Y_{\ell m_\ell}(\Omega)\times \chi_{1/2,m_s}]^{(j)}_m \label{spwfn} 
\end{equation}
The combined tensor product of these two is given by
\begin{equation}
\psi_{JM}(\vec{r}_1,\vec{r}_2)=[\phi_{\ell_1,j_1,m_1}(\vec r_1,{E_C}_1) \times \phi_{\ell_2,j_2,m_2}(\vec r_2, {E_C}_2)]^{(J)}_M\label{tpwfn}
\end{equation}
In $LS$-coupling for $\ell_1\neq \ell_2$ the antisymmetric wavefunction $\psi\left(\ell_1\ell_2 SLJM\right)$ is given by
\bea
\psi\left(\ell_{1}\ell_{2} SLJM\right)=\dfrac{1}{\sqrt{2}}\sum_{M_{S},M_{L}}\langle SM_SLM_L|SLJM\rangle \times
[\phi_{12}(\ell_{1}\ell_{2}LM_L)\chi_{12}(s_{1}s_{2}SM_S)- \phi_{21}(\ell_{2}\ell_{1}LM_L)\chi_{21}(s_{2}s_{1}SM_S)] \label{awf}
\eea
The matrix elements due to mutual interaction $V_{12}$ in $LS$-coupling of two particles are given by
\begin{eqnarray}
\langle\ell_a\ell_bSLJM|V_{12}|\ell_c\ell_dS'L'J'M'\rangle=
\sum\langle SM_SLM_L|SLJM\rangle\langle S'M'_SL'M'_L|S'L'J'M'\rangle \nonumber \\ 
\langle s_1m_{s_1}s_2m_{s_2}|s_1s_2SM_S\rangle\langle s'_1m'_{s_1}s'_2m'_{s_2}|s'_1s'_2S'M'_S\rangle  
\langle \ell_am_a\ell_bm_b|\ell_a\ell_bLM_L\rangle \langle \ell_cm_c\ell_dm_d|\ell_c\ell_dL'M'_L\rangle \nonumber \\
\sum_{\ell m}(-1)^{2(\ell-m)}
\begin{pmatrix}
  \ell & \ell_a & \ell_b \\
  -m & m_a & m_b 
\end{pmatrix}
\begin{pmatrix}
  \ell & \ell_c & \ell_d \\
  -m & m_c & m_d 
\end{pmatrix}
\langle \ell\|Y_{\ell_a}\|\ell_b\rangle^*\langle \ell'\|Y_{\ell_c}\|\ell_d\rangle 
\int R_{n_a\ell_a}^*\left(r\right)R_{n_b\ell_b}^*\left(r\right)\frac{1}{r^2}R_{n_c\ell_c}\left(r\right)R_{n_d\ell_d}\left(r\right) dr \label{me12}
\end{eqnarray}
An attractive pairing contact delta interaction has been used, $V_{12}=-g\delta(\vec r_1 - \vec r_2)$ for simplicity, because we 
can reach the goal with only one parameter adjustment.

\section{Results}
The major ingredients for the complete study of $^6$He are the matrix elements of pairing interaction.
These correspond to the radial integrals and to the coefficients (calculated in Mathematica notebook mentioned in block $3$ of procedure). 
For ground state these coefficients are used for calculating the contribution of various configurations and some ground state properties (see Table $2$ and Table $3$ of \cite{JSi}).
The coefficients of the matrix elements of Eq. (\ref{me12}) for $0^+, 1^-, 2^+$ and $3^-$ are summarized in Tables $1$ to $4$ below. 
These coefficients will lead to the calculation of higher excited states in low lying 
continuum of $^6$He. For all these states the coefficients of matrix elements are calculated for upper diagonal part of the matrix only due to symmetry.
\begin{table}[h]
\centering
\caption{Coefficients of ground and continuum ($0^+$) states of $^{6}$He.}
\vspace*{0.3cm}
\begin{tabular}{|l|l|l|l|l|}
\hline
$s_{1}s_{1}-s_{1}s_{1}$ & $s_{1}s_{1}-p_{1}p_{1}$ & $s_{1}s_{1}-p_{3}p_{3}$ & $s_{1}s_{1}-d_{3}d_{3}$ & $s_{1}s_{1}-d_{5}d_{5}$ \\
$1/2\pi$                  & $-1/2\pi$                 & $-1/\sqrt{2}\pi$          & $1/\sqrt{2}\pi$           & $\sqrt{3}/2\pi$           \\ \hline
\multirow{8}{*}{}         & $p_{1}p_{1}-p_{1}p_{1}$ & $p_{1}p_{1}-p_{3}p_{3}$ & $p_{1}p_{1}-d_{3}d_{3}$ & $p_{1}p_{1}-d_{5}d_{5}$ \\
                          & $1/2\pi$                  & $1/\sqrt{2}\pi$           & $-1/\sqrt{2}\pi$          & $-\sqrt{3}/2\pi$          \\ \cline{2-5} 
                          & \multirow{6}{*}{}         & $p_{3}p_{3}-p_{3}p_{3}$ & $p_{3}p_{3}-d_{3}d_{3}$ & $p_{3}p_{3}-d_{5}d_{5}$ \\
                          &                           & $1/\pi$                   & $-1/\pi$                  & $-\sqrt{3/2}/\pi$         \\ \cline{3-5} 
                          &                           & \multirow{4}{*}{}         & $d_{3}d_{3}-d_{3}d_{3}$ & $d_{3}d_{3}-d_{5}d_{5}$ \\
                          &                           &                           & $1/\pi$                   & $\sqrt{3/2}/\pi$          \\ \cline{4-5} 
                          &                           &                           & \multirow{2}{*}{}         & $d_{5}d_{5}-d_{5}d_{5}$ \\
                          &                           &                           &                           & $3/2\pi$                  \\ \hline
\end{tabular}
\end{table}
\begin{table}[h]
\centering
\caption{Coefficients of $1^-$ states of $^{6}$He.}
\vspace*{0.3cm}
\begin{tabular}{|l|l|l|l|l|}
\hline
$s_{1}p_{1}-s_{1}p_{1}$ & $s_{1}p_{1}-s_{1}p_{3}$ & $s_{1}p_{1}-p_{1}d_{3}$ & $s_{1}p_{1}-p_{3}d_{3}$ & $s_{1}p_{1}-p_{3}d_{5}$ \\
$1/6\pi$                  & $1/3\sqrt{2}\pi$          & $-1/3\sqrt{2}\pi$         & $2/3\sqrt{10}\pi$         & $-1/\sqrt{10}\pi$         \\ \hline
\multirow{8}{*}{}         & $s_{1}p_{3}-s_{1}p_{3}$ & $s_{1}p_{3}-p_{1}d_{3}$ & $s_{1}p_{3}-p_{3}d_{3}$ & $s_{1}p_{3}-p_{3}d_{5}$ \\
                          & $1/3\pi$                  & $-1/3\pi$                 & $1/3\sqrt{5}\pi$          & $-1/\sqrt{5}\pi$          \\ \cline{2-5} 
                          & \multirow{6}{*}{}         & $p_{1}d_{3}-p_{1}d_{3}$ & $p_{1}d_{3}-p_{3}d_{3}$ & $p_{1}d_{3}-p_{3}d_{5}$ \\
                          &                           & $1/3\pi$                  & $-1/3\sqrt{5}\pi$         & $1/\sqrt{5}\pi$           \\ \cline{3-5} 
                          &                           & \multirow{4}{*}{}         & $p_{3}d_{3}-p_{3}d_{3}$ & $p_{3}d_{3}-p_{3}d_{5}$ \\
                          &                           &                           & $1/15\pi$                 & $-1/5\pi$                 \\ \cline{4-5} 
                          &                           &                           & \multirow{2}{*}{}         & $p_{3}d_{5}-p_{3}d_{5}$ \\
                          &                           &                           &                           & $3/2\pi$                  \\ \hline
\end{tabular}
\end{table}

\begin{table}[]
\centering
\caption{Coefficients of $3^-$ states of $^{6}$He.}
\label{my-label}
\begin{tabular}{|l|c|l|}
\hline
$p_{1}d_{5}-p_{1}d_{5}$       & $p_{1}d_{5}-p_{3}d_{3}$ & $p_{1}d_{5}-p_{3}d_{5}$               \\
\multicolumn{1}{|c|}{$3/14\pi$} & $-3\sqrt{3/10}/7\pi$      & \multicolumn{1}{c|}{$3/7\sqrt{5}\pi$}   \\ \hline
\multirow{4}{*}{}               & $p_{3}d_{3}-p_{3}d_{3}$ & $p_{3}d_{3}-p_{3}d_{5}$               \\
                                & $9/35\pi$                 & \multicolumn{1}{c|}{$-3\sqrt{6}/35\pi$} \\ \cline{2-3} 
                                & \multirow{2}{*}{}         & $p_{3}d_{5}-p_{3}d_{5}$               \\
                                &                           & \multicolumn{1}{c|}{$6/35\pi$}          \\ \hline
\end{tabular}
\end{table}

\section{Conclusion}
We have outlined the method and the formulation that we used to calculate $^6$He states based on a continuum basis of $^5$He.
We have discussed the computational procedure in some detail and we have tabulated the coefficients that are needed in the calculation of matrix elements of the pairing interaction.
We intend to use these coefficients to study the electromagnetic response of $^6$He for transitions to the continuum (monopole, dipole, quadrupole etc. strength distributions), where the 
separate contribution of different configurations will be evaluated. This will allow us to make predictions on the continuum spectrum of this nucleus.

\section{Acknowledgements}
We would like to thank A.Vitturi, R.Chatterjee, J.A.Lay and Sukhjeet Singh for useful suggestions. J.Singh gratefully acknowledges the financial support from Fondazione Cassa di Risparmio di Padova e Rovigo (CARIPARO).

\begin{sidewaystable}[h]
\centering
\caption{Coefficients of $2^+$ states of $^{6}$He.}
\vspace*{0.3cm}
\begin{tabular}{|l|l|l|l|l|l|l|}
\hline
$s_{1}d_{3}-s_{1}d_{3}$ & $s_{1}d_{3}-s_{1}d_{5}$ & $s_{1}d_{3}-p_{1}p_{3}$ & $s_{1}d_{3}-p_{3}p_{3}$ & $s_{1}d_{3}-d_{3}d_{3}$ & $s_{1}-d_{3}d_{3}d_{5}$ & $s_{1}d_{3}-d_{5}d_{5}$    \\
$1/6\pi$                & $-\sqrt{3/2}/5\pi$      & $-1/5\pi$               & $-1/5\pi$               & $1/5\pi$                & $\sqrt{3/7}/5\pi$      & $2\sqrt{3/7}/5\pi$         \\ \hline
\multirow{12}{*}{}     & $s_{1}d_{5}-s_{1}d_{5}$ & $s_{1}d_{5}-p_{1}p_{3}$ & $s_{1}d_{5}-p_{3}p_{3}$ & $s_{1}d_{5}-d_{3}d_{3}$ & $s_{1}d_{5}-d_{3}d_{5}$ & $s_{1}d_{5}-d_{5}d_{5}$    \\
                       & $3/10\pi$              & $\sqrt{3/2}/5\pi$      & $\sqrt{3/2}/5\pi$      & $-\sqrt{3/2}/5\pi$     & $-3/5\sqrt{14}\pi$     & $-3\sqrt{2/7}/5\pi$       \\ \cline{2-7} 
                       & \multirow{10}{*}{}     & $p_{1}p_{3}-p_{1}p_{3}$ & $p_{1}p_{3}-p_{3}p_{3}$ & $p_{1}p_{3}-d_{3}d_{3}$ & $p_{1}p_{3}-d_{3}d_{5}$ & $p_{1}p_{3}-d_{5}d_{5}$    \\
                       &                        & $1/5\pi$               & $1/5\pi$               & $-1/5\pi$              & $-\sqrt{3/7}/5\pi$     & $-2\sqrt{3/7}/5\pi$       \\ \cline{3-7} 
                       &                        & \multirow{8}{*}{}      & $p_{3}p_{3}-p_{3}p_{3}$ & $p_{3}p_{3}-d_{3}d_{3}$ & $p_{3}p_{3}-d_{3}d_{5}$ & $p_{3}p_{3}-d_{5}d_{5}$    \\
                       &                        &                        & $1/5\pi$               & $-1/\pi$               & $-\sqrt{3/7}/5\pi$     & $-2\sqrt{3/7}/5\pi$       \\ \cline{4-7} 
                       &                        &                        & \multirow{6}{*}{}      & $d_{3}d_{3}-d_{3}d_{3}$ & $d_{3}d_{3}-d_{3}d_{5}$ & $d_{3}d_{3}-d_{5}d_{5}$    \\
                       &                        &                        &                        & $1/5\pi$               & $\sqrt{3/7}/5\pi$      & $2\sqrt{3/7}/5\pi$        \\ \cline{5-7} 
                       &                        &                        &                        & \multirow{4}{*}{}      & $d_{3}d_{5}-d_{3}d_{5}$ & $d_{3}d_{5}-d_{5}d_{5}$    \\
                       &                        &                        &                        &                        & $3/35\pi$              & $6/35\pi$                 \\ \cline{6-7} 
                       &                        &                        &                        &                        & \multirow{2}{*}{}      & $s_{1}d_{3}-d_{5}d_{5}$ \\
                       &                        &                        &                        &                        &                        & $12/35\pi$                 \\ \hline
\end{tabular}
\end{sidewaystable}

\end{document}